\documentclass[proceedings, preprint]{rmaa}

\usepackage{paralist}

\usepackage{psfrag,color}

\newcommand{\pasa}{Publications of the Astronomical Society of Australia}
  
\SetYear{2023}
\SetConfTitle{Astrorob 2023}

\title{The impact of the site quality of the Zadko Observatory on its scientific results} 

\author{
  Bruce Gendre,\altaffilmark{1,2,3} 
	Richard Tonello,\altaffilmark{4,5} 
	Mitchell Studdert,\altaffilmark{4,5} 
	David Coward,\altaffilmark{1} 
	Alain Klotz,\altaffilmark{6} 
	Eloise Moore,\altaffilmark{1} 
	John Moore,\altaffilmark{1} 
	and Fiona Panther\altaffilmark{1,7}}

\altaffiltext{1}{University of Western Australia, OzGrav ARC Centre of Excelence, 35 Stirling Highway, M013, 6009 Crawley, WA, Australia.}
\altaffiltext{2}{Present address: University of the Virgin Islands, 2 John Brewer's Bay, St Thomas, 00802, VI, USA.} 
\altaffiltext{3}{Email: bruce.gendre@gmail.com}
\altaffiltext{4}{Gravity Discovery Center, 1098 Military Road, 6503 Yeal, WA, Australia.}
\altaffiltext{5}{Chiro Observatory, IAU code 320.}
\altaffiltext{6}{IRAP, Universite de Toulouse, CNRS, CNES, UPS, 31401, Toulouse, France.}
\altaffiltext{7}{Forest Fellow.}

\shortauthor{Gendre, et al.}
\shorttitle{Zadko Observatory Science}

\listofauthors{B. Gendre, R. Tonello, M. Studdert, D. Coward, A. Klotz, E. Moore, J. Moore, F. Panther}
\indexauthor{Gendre, B.}
\indexauthor{Tonello, R.}
\indexauthor{Studdert, M.}
\indexauthor{Coward, D.}
\indexauthor{Klotz, A.}
\indexauthor{Moore, E.}
\indexauthor{Moore, J.}
\indexauthor{Panther, F.}

\abstract{We present the site quality of the Zadko Observatory, focusing on the brightness of the sky and the effects of the weather on the efficiency of the Observatory. We discuss these effects and their consequences for the scientific goals of the Observatory. Without surprise, the overall sky quality had decreased during the last decade. However, this decrease is mostly due to the weather pattern at the Observatory rather than the light pollution. We put an emphasis on how important the preservation of the sky darkness is for small observatories, in order to stabilize the global degradation of the site quality, as this directly impacts the scientific  return of the observations.}

\resumen{Presentamos el estudio de la calidad del cielo del Observatorio Zadko, centr\'andonos en el brillo del cielo y los efectos del clima en la eficiencia del observatorio. Discutimos estos efectos y sus consecuencias para los objetivos cient\'ificos del observatorio. Como se esperaba, la calidad general del cielo ha disminuido durante la \'ultima d\'ecada. Sin embargo, esta disminuci\'on se debe principalmente al patr\'on clim\'atico en el emplazamiento m\'as que a la contaminaci\'n lum\'inica. Ponemos \'enfasis en que lo importante es la preservaci\'on de la oscuridad del cielo para los observatorios peque\~nos, con el fin de estabilizar la degradaci\'on global de la calidad del sitio, ya que esto impacta directamente el retorno cient\'ifico de las observaciones.}

\addkeyword{History and philosophy of astronomy}
\addkeyword{Site testing}
\addkeyword{Atmospheric effects}
\addkeyword{Light pollution}

\begin{document}
\maketitle

\section{Introduction}
\label{sec:intro}

The Zadko Observatory is located approximately 70 kilometres north of Perth in the Shire of Gingin, Western Australia \citep{cow17, moo21}. It hosts the 1.0 m Ritchey‐Chrétien f/4 Zadko Telescope, which was installed in June 2008; this telescope is the only fully robotic telescope hosted by a Western Australian University (UWA), and is the only metre‐class optical facility at this southern latitude between the east coast of Australia and South Africa. At the time of its construction, the Zadko Observatory was a state-of-the-art facility, and for many years the Zadko Telescope performed observation campaigns in collaboration with various Australian and international research teams \citep{cow11}. This is mostly due to the Observatory being in a very desirable location: some parts of the sky can be observed only from the west coast of Australia in night conditions in the early morning. This has provided a niche opportunity for engaging in pure research collaborations, international partnerships and research training on site \citep{laa11}, with approximately one hundred scientists both National and International having worked with the Observatory. Since that time, the observatory has expanded to host several other instruments, which are devoted to various research or commercial programs.

In a companion paper (Moore et al., these proceedings), we have described how the global operations of the observatory are managed. In this paper, instead, we will focus on the scientific goals of the observatory, and how the quality of the site impacts those goals. In Section \ref{sec:goal} we will explain the scientific goals of the Observatory. In Section \ref{sec:dark} we will explain how the darkness of the sky is impacting our goals and how we try to mitigate this effect. In Section \ref{sec:bodies} we show the effect of celestial bodies on the observations. In Section \ref{sec:weather} we will describe in more detail the effects of the weather on the assets of the Observatory, before concluding. A forthcoming paper (Gendre et al., in preparation) will explore in more detail the weather patterns at the Observatory which are not addressed here.

\section{Scientific goals of the observatory}
\label{sec:goal}

Historically, the Zadko Observatory has been used to study the transient sky \citep{cow10}, and particularly explosive events, such as gamma-ray bursts, supernovae, and the emitters of gravitational wave sources \citep{aba12,and17,ant20, gen21}. However, due to the transient and random nature of these events, the Observatory was using fillers such as asteroid observations, long term monitoring of red dwarf stars, satellite follow-up, or even space junk monitoring \citep{zik20, mic21}.

However, Australia is expanding its role in the space sector, both commercially and in strategic defence. Space situational awareness programs, for instance, are an increasing priority. Space debris can be a serious hazard for commercial, communications, military and research satellites. 

Potential Earth colliders, so-called hazardous Near-Earth Objects (NEOs) \citep{gla00} are another well-known threat and are usually detected only a few hours before impact, and can pose a direct hazard to human activities. Space agencies, such as ESA, monitor the sky continuously to provide an early warning for such an event, of which the Zadko Telescope has been utilized for this task in the preceding few years \citep{con21}. All of those programs are now routinely run at the Observatory, and with the current upgrade of the facility, it is foreseen they will take more and more importance.

One of those programs, for instance, uses the Zadko Telescope to perform follow-up of radio observation of asteroids (Kruzins et al., in preparation). This kind of program requests a long duration follow-up, as the celestial body is in rotation, with a stable sky transparency, and a low brightness of the sky, due to the objects being very faint. In the following, we will use this program as a template for our considerations.

\section{Impact of the dark sky and how to mitigate light pollution}
\label{sec:dark}

The Zadko Observatory has been established on a pre-existing scientific precinct, beside another Observatory aimed at outreach activities, the Gravity Discovery Center Observatory (hereafter the GDC observatory)\footnote{See https://gravitycentre.com.au/wp-content/uploads/2023/12/Annual-Report-2022-2023-Final-2-1.pdf for more information}. It was constructed in 2009 with, at that time, no controls to evaluate or mitigate the effects of light pollution. However, with the continued expansion of Perth’s northern suburbs it was noted there has been a significant increase in light pollution on the southern horizon \citep{fal16}. In fact, under favourable conditions of low humidity and no cloud the light dome extends to an altitude of approximately 25 degrees above the southern horizon. In the case of unfavourable conditions, the light pollution can reach an altitude of 35 degrees above the southern horizon.

Starting in 2018, astronomers at the GDC observatory, preoccupied by the increasing glow of the sky and its impact on the outreach activities, began monitoring of the sky brightness and management of onsite lighting to reduce the effects of local light pollution on both the Zadko and GDC observatories. For this purpose, they established the Gingin Dark Sky Reserve, which encompasses 3211 Km$^2$ of protected dark sky, centered around the GDC and Zadko Observatories.

\begin{figure}[!bth]
  \includegraphics[width=\columnwidth]{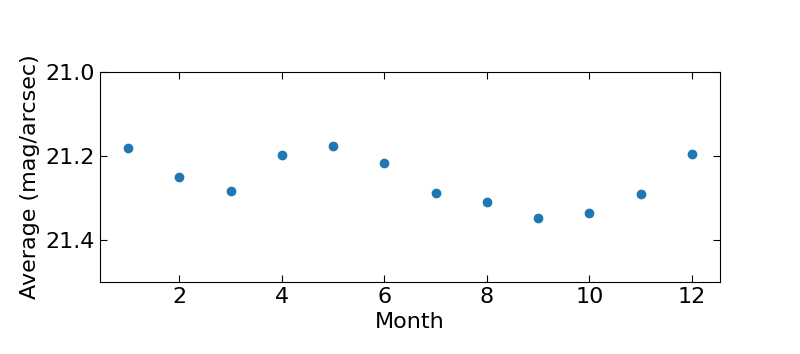}
  \caption{Evolution of the mean darkness of the sky along the year, with the effect of the Galactic bulge clearly visible. The measurements have been taken at the start of the night, explaining why the effect of the Galactic bulge is at its maximum a couple of months before it reaches the zenith.}
  \label{fig:sky}
\end{figure}

\begin{figure}[!bht]
  \includegraphics[width=\columnwidth]{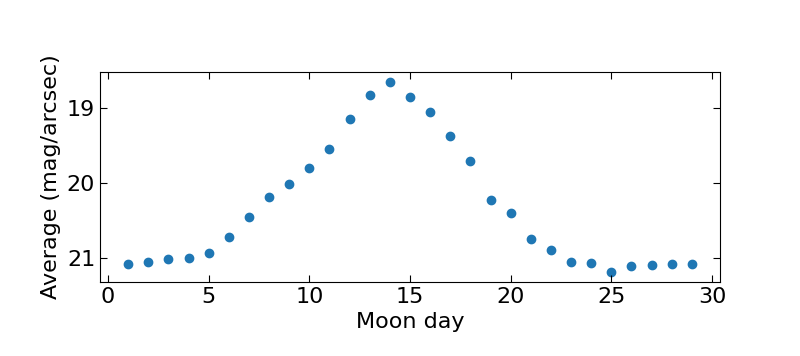}
  \caption{Breakout of the sky luminosity along one Moon cycle, averaged over 12 months. The new moon is located at phase 0.}
  \label{fig:moon}
\end{figure}

They started by replacing the majority of outdoor lighting on the site, and  brought them into compliance with international dark sky standards. The impact of that local light pollution on the observatories was however insignificant, as one could expect: most of the issues arises from the diffused light and not the direct sources \citep{han18}. In fact, observations taken in 2023 (Gendre et al. in preparation) have shown that with all onsite lighting turned on, the the limiting magnitude at zenith was reduced by only 0.1 magnitude.

The GDC astronomers also started to continually monitor the sky quality, in order to record any changes, establish trends, and assess whether local light pollution has increased/decreased within the reserve. For that purpose, a Unihedron SQM-LU sky quality instrument has been installed on the roof of the GDC observatory, together with an array of detectors within the reserve itself. These observations demonstrated that the light pollution dome from Perth is the largest factor in limiting magnitude \citep[as it can be also infered from][]{fal16}, with an effect of 0.5 magnitudes. 

With the dark sky reserve, centred around the Gingin Gravity Precinct to help mitigate the ingress of light pollution towards the observatories, and a strict lighting management plan in place within 40 kilometres around the observatories, the sky quality started to stabilize. With all those efforts, the current sky brightness is at mimimum 21.4 magnitude.arcsec$^{-1}$, which gives the limit of what can be observed on site.

\section{Effect of celestial bodies on our scientific goals}
\label{sec:bodies}

Natural celestial bodies can also affect the sky background \citep{fal16, han18}. While this is barely noticeable within cities, in a modern observatory, established at a reasonable distance from light pollution, the sky transparency allows one to notice the effect of the Galaxy bulge, which rises in the sky in April, and passes close to the zenith in June-July in the Southern hemisphere. In Figure \ref{fig:sky}, it is possible to see that the Galaxy has a 0.2 magnitude effect on the sky luminosity, and that the best period to observe faint objects is during the Southern Hemisphere's Spring.

Another effect clearly at play is the effect of the Moon, as anybody could guess. We present it on Figure \ref{fig:moon}. Following the standard naming convention (clear nights being the nights close to the full moon, and dark ones being close to the new moon), clear nights are about 3 magnitudes brighter than dark nights, and this has some obvious consequences on the faintest objects one can detect with the Zadko telescope: during clear nights with full moon, it is very difficult to detect objects fainter than about 18.5 magnitudes.

This has some consequences, as the asteroids observed for the current program are ranging in luminosity between 12 and 20. While there is no restrictions for bright objects, this can bias our observation program toward nearby or large events (i.e. those which are reflecting more light and/or presenting less attenuation due to the distance). For this reason, it is clear that it is impossible to dedicate the Zadko telescope to a single class of event, and that the scientific scheduling of the observations has to be carefully planned.

\section{Impact of the weather and how to protect the assets}
\label{sec:weather}

\begin{figure}[!htb]
  \includegraphics[width=\columnwidth]{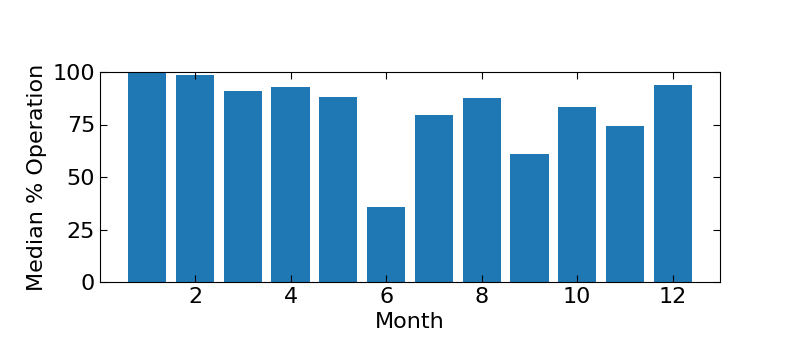}
  \caption{Mean efficiency of the Zadko Observatory during a typical year, averaged over one month, and using the same method than in Figure \ref{fig:eff}. One can clearly see the effect of the wet season.}
  \label{fig:year}
\end{figure}

The second important effect one has to consider is that of the weather, particularly the rain and the humidity \citep{abb21}. It is obviously not desirable to have the roof of an observatory open whilst it is raining, which in turn can have a serious impact on available viewing time. As most modern observatories use electronic cameras which have a given working range for temperature, humidity and pressure, it is the responsibility of the observatory operators to maintain those parameters in the correct range to protect these valuable assets. A normal operation efficiency is of the order of 60\% for a professional observatory not located on a top quality site \citep{abb21}. In Figure \ref{fig:year} we present the efficiency of the site for the year 2023. 

\begin{figure*}[!tb]
  \includegraphics[width=\textwidth]{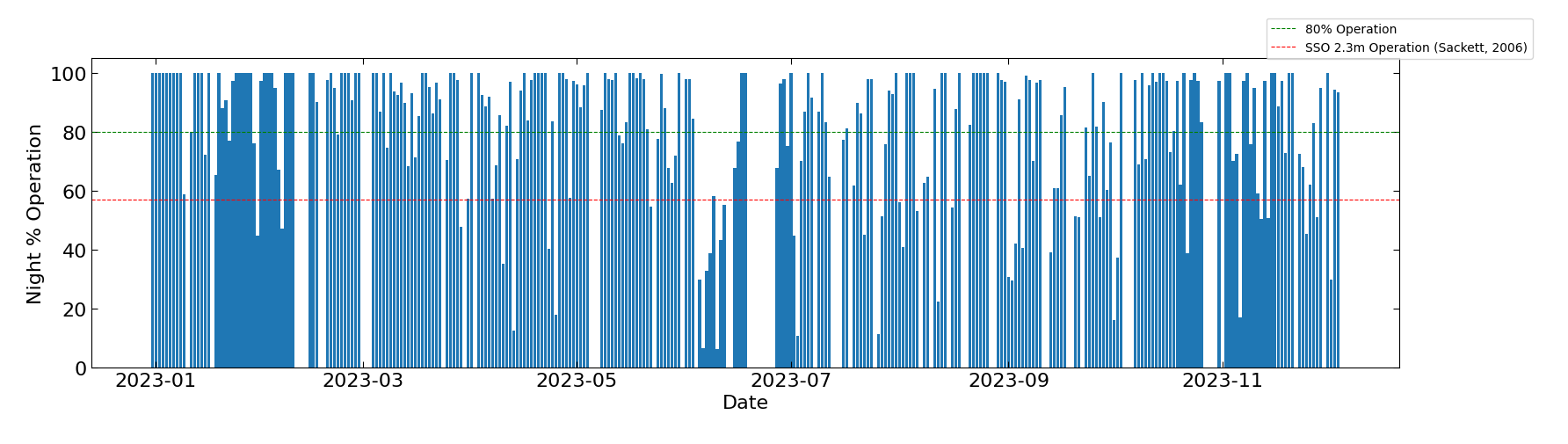}
  \caption{Efficiency of the Zadko Observatory, expressed as the amount of time the roof was opened vs the amount of available time during a given night. The dotted red line represent the Siding Spring Observatory efficiency measured in 2006, as indicated in \citet{sac06}}
  \label{fig:eff}
\end{figure*}

Our overall operation efficiency is of the order of 80\%, which is very good for an observatory like the Zadko Observatory, with some months being near an efficiency of 100\%. Those nights are mainly during the hot, dry season (traditionally known as "Birak" by the Noongar People). This is confirmed with more granularity (as displayed in Figure \ref{fig:eff}): for a fair amount of nights within the year we reach an efficiency of 100\%. However, one can clearly see that most nights have a period of some minutes to a couple of hours of closure, due to the humidity coming from the sea. In winter there are many rainy and foggy nights preventing operation during the whole night.

This has some consequences on how long a continuous observation can be. While during the summer it is possible to have long observations, this is not the case during the winter. Indeed, one effect not presented on these figures is the rapid evolution of the seeing of the site (which can change dramatically within minutes), indicating that there are cloud sheets passing through the sky (see Gendre et al. for a full discussion of this effect). Each of them can trigger the closure of the roof to protect the observatory assets, blocking a continuous observation. One can consider that observations undertaken at the Zadko Observatory need to factor in that they won't be continuous, except during the good season. This has consequences for all observation policies, and requests specific methods of data analysis to correct its effects within the data.

\section{Conclusion}

The Gingin Gravity Precinct, despite it being close to a major source of Light Pollution (Perth’s northern suburbs), was the logical place to locate the Zadko Observatory. With its established infrastructure, education and research programs, the Precinct is the perfect place to instigate, demonstrate and educate people. However, this has not taken into account the scientific goals set for an Observatory. Without constant evaluation and monitoring of the site parameters, it is clear from the recent data collected that the scientific astronomical activities of the Zadko Observatory may be at risk.

The current measures to continue to improve the darkness of the sky include that all external light fixtures are to be shielded or mounted underneath the awning of buildings to mitigate the amount of stray light scattered up into the air, and to have a colour corrected temperature of 3000K or less. At greater distance from the Observatory, a buffer zone of 3211 square kilometres had been created, in which the GDC astronomers are working with the local government to change public lighting to be dark sky compliant. Finally, a public education campaign is also in place to educate the locals on the effects of light pollution which can assist in mitigating its locally.

Albeit the dark sky reserve project was initiated to protect:
\begin{itemize}
\item	Astronomical Education and Tourism;
\item	Astronomical/ Astrophysical Research;
\item	Local Aboriginal Cultural Heritage;
\item	Local Wildlife, Bird, Insect, amphibian and Coastal marine life and habitats found within the local ecosystem
\end{itemize}
it has been of great help to mitigate the degradation of the dark sky quality. Future monitoring with upgraded equipment will give a better sample of the effect of local and distant light pollution domes and their impact. It will also assist in the annual reporting required by Dark Sky International (formerly International Dark Sky Association) to demonstrate the continual protection of the accredited Dark Sky Reserve.

Unfortunately, the climate change within the last few years has had some more serious effects, with some years being particularly humid, and an increasing trend of rainy nights preventing the observatory to stay operating the whole night. It is not possible to really mitigate this effect, but rather to take it into account when defining the future priorities and scientific goals of the observatory: the scientific policy of any observatory has to be revised from time to time to reflect changes in local conditions. Not doing so will only bring frustrations when the previous goals are not fulfilled or no longer attainable.

\section*{ACKNOWLEDGEMENTS}
This research was supported by the Australian Research Council Centre of Excellence for Gravitational Wave Discovery (OzGrav), through project number CE170100004. E.M. acknowledges support support from the Zadko Postgraduate Fellowship and International Space Centre (ISC). 

At the Gravity Precinct, located in the Shire of Gingin, the Gravity Discovery Centre \&
Observatory, and the University of Western Australia acknowledge the Yued Noongar people as the traditional owners of the land
on which it is situated. The Yued Noongar remain the spiritual and cultural custodians of this
land, where they continue to practise their values, languages, beliefs and knowledge.


\begin{thebibliography}
\bibitem[Abbot et al.(2021)]{abb21} Abbot, H.~J., Munro, J., Travouillon, T., et al.\ 2021, \pasp, 133, 095001. doi:10.1088/1538-3873/ac1f3b
\bibitem[Antier et al.(2020)]{ant20} Antier, S., Agayeva, S., Aivazyan, V., et al.\ 2020, \mnras, 492, 3904. doi:10.1093/mnras/stz3142
\bibitem[Andreoni et al.(2017)]{and17} Andreoni, I., Ackley, K., Cooke, J., et al.\ 2017, \pasa, 34, e069. doi:10.1017/pasa.2017.65
\bibitem[LIGO Scientific Collaboration et al.(2012)]{aba12} LIGO Scientific Collaboration, Virgo Collaboration, Abadie, J., et al.\ 2012, \aap, 539, A124. doi:10.1051/0004-6361/201118219
\bibitem[Conversi et al.(2021)]{con21} Conversi, L., Koschny, D., Micheli, M., Kreken, R., Moreta, P.~R., Kugel, U., Doelling, E., Cano, J.~L., Cennamo, R., Faggioli, L., Foglietta, A., Moissl, R., Oliviero, D., Petrescu, E., and Rudawska, R. 2021, “ESA's NEO Coordination Centre Observational Network,” in 7th IAA Planetary Defense Conference, Vienna
\bibitem[Coward et al.(2010)]{cow10} Coward, D.~M., Todd, M., Vaalsta, T.~P., et al.\ 2010, \pasa, 27, 331. doi:10.1071/AS09078
\bibitem[Coward et al.(2011)]{cow11} Coward, D.~M., Gendre, B., Sutton, P.~J., and Howell, E.~J. 2011, \mnras, 415, L26
\bibitem[Coward et al.(2017)]{cow17} Coward, D.~M., Gendre, B., Tanga, P., Turpin, D., Zadko, J., Dodson, R., Devogéle, M., Howell, E.~J. , Kennewell, J.~K., Boer, M., Klotz, A., Dornic, D., Moore, J.~A., and Heary, A. 2017, \pasa, 34, 16
\bibitem[Falchi et al.(2016)]{fal16} Falchi, F., Cinzano, P., Duriscoe, D., et al.\ 2016, Science Advances, 2, e1600377. doi:10.1126/sciadv.1600377
\bibitem[Gendre et al.(2021)]{gen21} Gendre, B., Coward, D., Moore, J., et al.\ 2021, Revista Mexicana de Astronomia y Astrofisica Conference Series, 53, 124. doi:10.22201/ia.14052059p.2021.53.24
\bibitem[Gladman et al.(2000)]{gla00} Gladman, B., Michel, P., and Froeschlé, C. 2000, Icarus, 146, 176
\bibitem[H{\"a}nel et al.(2018)]{han18} H{\"a}nel, A., Posch, T., Ribas, S.~J., et al.\ 2018, \jqsrt, 205, 278. doi:10.1016/j.jqsrt.2017.09.008
\bibitem[Laas-Bourez et al.(2011)]{laa11} Laas Bourez, M., Coward, D.~M., Klotz, A., Boer, M. 2011, Advances in Space Research, 47, 3
\bibitem[Micheli et al.(2021)]{mic21} Micheli, M., Koschny, D., Conversi, L., et al.\ 2021, Acta Astronautica, 184, 251. doi:10.1016/j.actaastro.2021.04.022
\bibitem[Moore et al.(2021)]{moo21} Moore, J.~A., Gendre, B., Coward, D.~M., et al.\ 2021, Revista Mexicana de Astronomia y Astrofisica Conference Series, 53, 35. doi:10.22201/ia.14052059p.2021.53.08
\bibitem[Sackett(2006)]{sac06} Sackett, P.~D. 2006, Report to the Australian Academy of Science, available at https://www.atnf.csiro.au/nca/sso.pdf
\bibitem[Zic et al.(2020)]{zik20} Zic, A., Murphy, T., Lynch, C., et al.\ 2020, \apj, 905, 23. doi:10.3847/1538-4357/abca90

\end{thebibliography}
\end{document}